 \documentclass{emulateapj}
 \usepackage{apjfonts}
 \usepackage{graphicx}
 \usepackage{psfig}
\usepackage{color}

\def\MgII{Mg\,{\sc ii}}

\def\d{\mathrm{d}}

\newcommand{\begit}{\begin{itemize}}
\newcommand{\enit}{\end{itemize}}
\newcommand{\begen}{\begin{enumerate}}
\newcommand{\enen}{\end{enumerate}}

\newcommand \be {\begin{equation}}
\newcommand \ee {\end{equation}}
\newcommand \bea {\begin{eqnarray}}
\newcommand \eea {\end{eqnarray}}

\newcommand \kms {\,{\rm km \,\, s}^{-1}}
\newcommand \cm {\,{\rm cm }}
\newcommand \pc {\,{\rm pc }}
\newcommand \yr {\,{\rm yr }}
\newcommand \s {\,{\rm s }}

\newcommand \g {\,{\rm g }}
\newcommand \kpc {\,{\rm kpc }}
\newcommand \K {\,{\rm K }}
\newcommand \M {\,{\cal M }}

\newcommand \Myr {\,{\rm Myr} }
\newcommand \rcl {r_{\rm cl}}
\newcommand \mcl {M_{\rm cl}}
\newcommand \vcl {v_{\rm ej}}

\newcommand \rc {{a}}

\newcommand \Rg {r_{\rm G}}
\newcommand \Mg {M_{\rm G}}
\newcommand \Msh {M_{\rm sh}}
\newcommand \Lcl {L_{\rm cl}}
\newcommand \Rd {R_{\rm d}}
\newcommand \eg {\epsilon_{\rm G}} 
\newcommand   \sfr {{\dot M_*}}
\newcommand   \hot {{\dot M_{\rm h}}}

\newcommand  \tcc {{t_{\rm cc}}}
\newcommand  \td {{t_{\rm dest}}}
\newcommand{\beqa}{\begin{eqnarray}}
\newcommand{\eeqa}{\end{eqnarray}}
\begin{document}

\title
[Star clusters drive super-galactic winds]
{Radiation pressure from massive star clusters\\ as a launching mechanism for super-galactic winds}

\author{
Norman Murray\altaffilmark{1,2},~
Brice M\'enard\altaffilmark{1}~
\&~ Todd A.~Thompson\altaffilmark{3,4,5}
}

\altaffiltext{1}{Canadian Institute for Theoretical Astrophysics, 60 St. George Street, University of Toronto, Toronto, ON M5S 3H8, Canada; murraycita.utoronto.ca} 
\altaffiltext{2}{Canada Research Chair in Astrophysics}
\altaffiltext{3}{Department of Astronomy, The Ohio State University, 140 W 18th Ave., Columbus, OH 43210; thompson@astronomy.ohio-state.edu.}
\altaffiltext{4}{Center for Cosmology \& Astro-Particle Physics, The Ohio State University, 191 W. Woodruff Ave., Columbus, OH 43210}
\altaffiltext{5}{Alfred P.~Sloan Fellow}

\submitted{Submitted to ApJ, \today}


\begin{abstract}
Galactic outflows of low ionization, cool ($\sim10^4\K$) gas are ubiquitous in local starburst galaxies, and in the majority of galaxies at high redshift. How these cool outflows arise is still in question. Hot gas from supernovae has long been suspected as the primary driver, but this mechanism suffers from its tendency to destroy the cool gas as the latter  is accelerated. We propose a modification of the supernova scenario that overcomes this difficulty. 

Star formation is observed to take place in clusters; in a given galaxy, the bulk of the star formation is found in the $\sim 20$ most massive clusters. We show that, for $L^\star$ galaxies, the radiation pressure from clusters with $\mcl \gtrsim 10^6\,M_\odot$ is able to expel the surrounding gas at velocities in excess of the circular velocity of the disk galaxy. This cool gas can travel above the galactic disk in less than $2\Myr$, well before any supernovae erupt in the driving cluster. Once above the disk, the cool outflowing gas is exposed to radiation, and supernovae induced hot gas outflows, from other clusters in the disk, which in combination drive it to distances of several tens to hundreds of kiloparsecs. Because the radiatively driven clouds grow in size as they travel, and because the hot gas is more dilute at large distance, the clouds are less subject to destruction if they do eventually encounter hot gas. Therefore, unlike wind driven clouds, radiatively driven clouds can survive to distances $\sim 50\kpc$, and potentially give rise to the metal absorbers seen in quasar spectra. We identify these cluster-driven winds with large-scale galactic outflows. In addition, we note that the maximum cluster mass in a galaxy is an increasing function of the galaxy's gas surface density. As a result, only starburst galaxies, where massive clusters reside, are able to drive winds cold outflows on galactic scales via this mechanism. We find that the critical star formation rates above which large scale cool outflows will be launched to be $\dot\Sigma^{crit}_*\approx 0.1\, M_\odot\yr^{-1}\kpc^{-2}$, which is in good agreement with observations.
\end{abstract}




\section{Introduction}


Observational studies of both nearby and high redshift star forming galaxies have established that cold gas emerges from such galaxies at velocities ranging from a few tens to several hundred kilometers per second \citep{1990ApJS...74..833H,1996ApJ...462L..17S,1997ApJ...486L..75F,2000ApJS..129..493H,2000ApJ...528...96P,2001ApJ...554..981P,2003ApJ...588...65S,2005ApJ...621..227M,2005ApJS..160..115R,2007ApJ...663L..77T,2009ApJ...692..187W,2009arXiv0912.3263M}. Hot outflowing gas from supernovae (SN) has been considered to be the primary mechanism for driving galactic outflows (\cite{1985Natur.317...44C,2002ASPC..254..292H} and references therein). It is generally assumed that cold gas and dust are entrained in the hot gas.

As numerous authors have pointed out, in such a scenario the cold gas is subject to Kelvin-Helmholtz instabilities \citep{1994ApJ...420..213K,2008ApJ...674..157C} and/or by conductive evaporation \citep{2005MNRAS.362..626M} and ends up surviving for less than $1\Myr$ (a few cloud crushing times) before being destroyed. The cloud crushing time follows $\tcc=(\rho_c/\rho_h)^{1/2}\,\rc/v_b$, where $\rho_c/\rho_h$ is the density ratio between the cold gas and the hot gas, $\rc$ is the radius of the cloud before it is exposed to the hot gas, and $v_b\approx1000\kms$ is the shock velocity in the hot gas. This implies that the cold gas can travel less than a few hundred parsecs before being mixed into the hot gas and hence effectively destroyed. However, observational evidence shows that cold gas survives at least out to ten $\kpc$. An explanation for the presence of this cold gas is required.
In this paper we show how taking into account the effect of radiation pressure on dust grains can affect the fate of the gas.

A second point regarding the driving of outflows is that most stars form in massive clusters, both in quiescent spirals that lack strong outflows, like the Milky Way, and in starburst galaxies such as M82. We argue that outflows emerge from the most massive clusters in a galaxy, as opposed to arising from the collective effects of smaller clusters. The radiation pressure from small clusters will not punch holes in the gas disk, and the subsequent supernovae will either leak into the interstellar medium (ISM) in low star formation rate galaxies, or will  radiate their energy away after their growth has been halted by the high pressure of the ISM in high star formation rate galaxies (such as Arp 220). In contrast, the radiation pressure from massive clusters will blow cold gas out of the disk, paving the way for the hot gas from SN to escape.

In this paper we present a model including both radiation pressure from massive clusters and ram pressure from supernovae. 
As recently pointed out by \cite{2009MNRAS.396L..90N}, we find that before any cluster stars explode as SNe, radiation pressure, acting alone, launches the cold gas surrounding a star cluster to heights of several hundred parsecs above the disk. As the cold cloud travels above the disk, it expands. This expansion, as well as a lower hot gas density at larger radii, significantly increase the cloud crushing time once it does encounter hot gas, allowing it to survive long enough to reach distances exceeding one hundred $\kpc$. Once above the disk, a cold gas cloud experiences a ram pressure force which is comparable to the radiation pressure force \citep{2005ApJ...618..569M}.

While it is clear that both driving mechanisms are significant, the action of the radiation pressure is crucial for the survival of cold gas out to large radii. We also illustrate the importance of this force by showing that it is capable of driving outflows even in the absence of hot gas. In \S \ref{sec:starclusters} we describe how the masses of the largest star clusters in a galaxy are determined by the interplay between self-gravity and radiation pressure in giant molecular clouds. In \S \ref{sec:winds} we describe how winds are launched from massive clusters, and how this leads to a critical star formation rate to produce galactic winds. We follow the evolution of these winds outside the star forming disk, to distance of order the galactic virial radius, in \S \ref{sec:galacticwinds}. We describe the cloud properties, cover factor, and mass loss rates of the wind in \S \ref{sec:properties}. We discuss our results in \S \ref{sec:discussion} and offer conclusions in the final section.

\section{PROPERTIES OF STAR CLUSTERS}\label{sec:starclusters}

We show below that the cluster mass and radius determine the dynamics of the interstellar medium (ISM) around the cluster, out to distances comparable to the disk scale height. We describe how to estimate the cluster mass as a function of the surface density of the parent giant molecular cloud. We then describe the (observationally determined) cluster radius as a function of cluster mass. Finally, we describe the cluster mass distribution function, and conclude that most of the luminosity in a starburst galaxy is typically produced by a dozen or so massive star clusters.

\subsection{Cluster masses}

Star clusters form in dense cores inside GMCs. As shown by \cite{2010ApJ...709..191M}, radiation from the young stars will transfer momentum to the surrounding gas through dust grains. The energy from the absorbed photons is then reemitted isotropically in the infrared; in most galaxies these IR photons simply escape the galaxy. 
\footnote{If the optical depth in the far-infrared, $\tau_{FIR}$, is larger than unity, which is the case in ultraluminous infrared galaxies (ULIRGs) and submillimeter galaxies, photons will experience multiple interactions with dust grains. This can enhance the effectiveness of radiation pressure by a factor of $\tau_{FIR}$.}
The bolometric luminosity of a young cluster is carried predominantly by ultraviolet radiation. The dust opacity in the ultraviolet is typically $\kappa \sim 1000\cm^2\g^{-1}$, so these ultraviolet photons see an absorption optical depth of order unity at a column of $N_H\sim 10^{21}\cm^{-2}$. Since the gas in the vicinity of a massive proto-cluster has a density in excess of $10^4\cm^{-3}$, the corresponding length scale is about $10^{17}\cm$, i.e. $0.03\pc$. As a result, $\tau_{UV}>1$. In the single-scattering regime (where $\tau_{FIR}<1$) essentially all photons emitted by the cluster are absorbed once and then leave the system as infrared photons. It follows that the radiation pressure force is
\begin{equation}
F_{\rm rad} \simeq {L\over c}\,.
\end{equation}

The accretion of gas and the formation of stars occur until the radiation pressure becomes comparable to the gravitational force acting on the infalling material: $F_{rad}\sim F_{grav}$. At this stage, the infall slows down significantly and likely takes place through a disk. 
By equating these two forces, \cite{2010ApJ...709..191M} showed that the fraction of gas $\epsilon_G\equiv \mcl/\Mg$ in a GMC that is converted into stars is proportional to the surface density $\Sigma_G$ of the parent GMC, for GMCs that are optically thin to far-infrared radiation,
\be
\epsilon_G={\pi G c\over 2(L/M_*)}\Sigma_G,
\label{eq:eps_g}
\ee
where $L/M_*$ is the light to mass ratio of the cluster. This relation shows that the characteristic (or maximum) mass of the star clusters in a galaxy will scale with the surface density of the galactic disk $\Sigma_g$ and hence the star formation rate of the host galaxy. We will make use of the above relation in section~\ref{sec:scaling}.

The star formation process is believed to be rather inefficient. Typical star formation rates in massive ($10^5-10^6M_\odot$) proto-clusters are $\sim1\,M_\odot\yr^{-1}$, so the time to double the stellar mass is a significant fraction of the dynamical time of the parent GMC, and the main sequence lifetime of a massive star. As long as the mass in stars $M_*<\eg \Mg$, the cluster does not strongly affect the GMC. However, once the stellar mass reaches this upper bound, the cluster begins to disrupt the cloud, eventually cutting off its own fuel supply. The disruption takes less than the GMC dynamical time \citep{2010ApJ...709..191M}, so that further star formation will only increase the cluster luminosity by a factor of a few at most. \emph{The radiation force exerted on the surrounding gas can therefore only be a few times larger than the gravitational force}. As a result, the ejection velocity with which radiation pressure can expel gas is of order the escape velocity of the system:
\begin{equation}
\vcl\sim v_{esc}=\sqrt{2G\mcl/\rcl}\,.
\label{eq:vc}
\end{equation}

\begin{figure}
\includegraphics[width=\hsize]{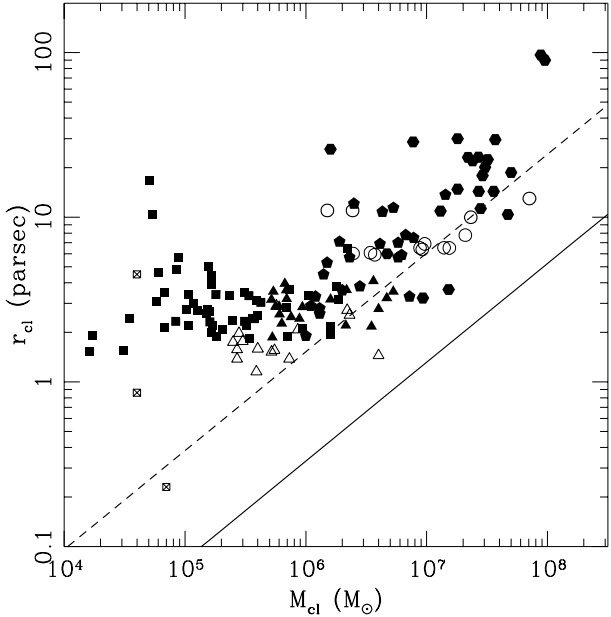}
\caption{The present day half light radii for star clusters. Filled squares depict Milky Way globular clusters, filled triangles M 31 globulars, open triangles correspond to M82 superclusters, filled pentagons are globular clusters from Cen A. Filled hexagons are ultra-compact dwarfs \citet{2005ApJ...627..203H,2007A&A...463..119H}}
\label{fig: sizes}
\end{figure}

\subsection{Size and mass distributions}
\label{sec:mass function}

Massive star clusters are compact. For example, Milky Way globular clusters have half light radii $r_h\sim2-3\pc$, independent of cluster mass, for clusters with masses between $10^4$ and $10^6M_\odot$ (see figure \ref{fig: sizes}). Observations of young clusters in other galaxies also find $r_h\sim2\pc$ independent of luminosity or $\mcl$ \citep{2007A&A...469..925S}. Younger clusters of the same mass tend to have even smaller half-light radii, e.g. \cite{2004A&A...416..537L} and \cite{2007ApJ...663..844M}. This is expected if the gas sloughed off by stars as they evolve is also ejected from the cluster; the lack of intercluster gas strongly implies that the latter is the case.

Clusters with $\mcl\gtrsim10^6M_\odot$ do show an increase in radius with increasing mass, of the form $\rcl\sim\mcl^{3/5}$ \citep{2005ApJ...627..203H,2009ApJ...691..946M}; see Figure \ref{fig: sizes}. Numerically, the velocity of a cluster is given by 
\bea
\vcl &\simeq& 100 \left( \frac{M_{\rm cl}}{10^6\,M_\odot} \right)^\alpha \kms
\eea
where $\alpha=1/2$ for $\mcl < 10^6 M_\odot$ and $\alpha=1/5$ for more massive systems. As discussed below, this quantity is crucial in determining whether a newly formed cluster can expel its surrounding material above a galactic disk and beyond.

The star cluster mass function is known to be shallow:
\be \label{eq: mass}
{dN\over d\mcl}\propto \mcl^{-\beta},
\ee
where $N$ is the number of clusters of mass $\mcl$, with $\beta\approx 1.8$. This suggests that a few to a few dozen clusters dominate the luminosity of a star forming galaxy, and hence certain aspects of the ISM dynamics. Direct counts of the number of clusters that provide half the star formation of the Milky Way \citep{2010ApJ...709..424M} and M82 \citep{2007ApJ...663..844M} (see their figure 8) are consistent with this conclusion.

\section{DRIVING WINDS FROM MASSIVE STAR CLUSTERS} \label{sec:winds}

\subsection{The disruption of giant molecular clouds} 

At early times, i.e. less than a few Myr after a massive star cluster forms, the forces acting on the surrounding material include only radiation pressure and shocked stellar winds. The first supernovae explode after about $4\Myr$.
Observations show that massive star clusters disrupt GMCs both in the Milky Way \citep{2010ApJ...709..424M} and in nearby galaxies, including the large Magellanic Cloud \citep{1996ApJ...465..231O} and the Antennae \citep{2007ApJ...668..168G}. As shown by \citet{2010ApJ...709..191M}, this disruption is primarily due to the effect of radiation pressure. 
While it has been thought that the dynamics of gas around star clusters was dominated by shocked stellar wind pressure, observations of HII regions in the Milky Way have shown that the pressure in such hot gas is equal to that of the associated HII ($10^4$K) gas \citep{1984ApJ...278L.115M,2009ApJ...693.1696H,2010ApJ...709..191M}, and therefore dynamically irrelevant.

Why don't shocked stellar winds affect the bubble dynamics? Examination of actively star forming regions in the Milky Way shows that the bubble walls are far from uniform; multiple shells, clumps, and 'elephant trunks' (pillars pointing toward the center of the bubble) are common \citep{2010ApJ...709..424M}. This is not unexpected, given the turbulent nature of the ISM; the gas near the cluster is clumpy, even before any star formation begins. The fact that the bubble walls do not actually isolate the bubble interior from the region outside the walls explains both the low luminosity of diffuse x-ray emission inside the bubbles, and the dynamical irrelevance of the hot gas \citep{2009ApJ...693.1696H}. However, it is worth stressing that there are clear shell structures surrounding the bubbles in {\em all} of the star forming regions examined by \citet{RM}, and that the bubble walls are moving radially outward at $10-20\kms$.

As outlined above, once the radiation pressure of a newly formed star cluster balances the gravitational force acting on the surrounding gas, the star formation rate slows. The luminosity gradually increases, driving gas out of the cluster.  As the system evolves, a bubble in the interstellar medium forms around the cluster. 

For our purpose, we can ignore the fact that the shells are clumpy. Since radiation pressure supplies the bulk of the momentum absorbed by the gas, the clumpy nature of the ISM will not prevent the formation of a bubble: any gas in the vicinity will absorb momentum from the radiation field. Some very dense clumps will not be appreciably accelerated, but the bulk of the gas in a GMC is in a relatively diffuse state ($n\sim100\cm^{-3}$) and will be pushed outward, piling up in a shell, or in multiple shells.

Over time the bubble expands, so the surrounding shell sweeps up more gas, and the shell mass $\Msh(r)$ increases with increasing radius. The evolution of the optical depth depends on the surface density through the shell, $\Sigma_{sh}\equiv \Msh(r)/4\pi r^2$; for a Larson-like GMC density profile $\rho(r)\propto r^{-1}$, $\Sigma_{sh}(r)$ is constant. When the shell, or some part of it, breaks out of the GMC, near the surface of the gas disk, the growth rate of $\Msh(r)$ slows and eventually halts. After that, the column of the shell or its fragments will decrease with increasing radius.

\cite{2010ApJ...709..191M} studied the $r<H$ behavior of such a shell in a GMC centered at the disk midplane. Their model included forces due to protostellar jets, shocked stellar winds, HII gas pressure, turbulent pressure, gravity and radiation pressure. The jets exert an outward force, but are relevant only on small scales $\sim 1\pc$; the shocked stellar winds may also be confined, briefly, on these scales; we will refer to the combination of the two as the small-scale force, $F_{sm}$. These authors found that, on larger scales, the radiation pressure and the gravity dominate the dynamics when the star cluster has $\mcl\gtrsim3\times10^4M_\odot$. 

Since the largest GMCs have radii only a factor of two or three smaller than the disk scale height, bubbles capable of disrupting their parent GMCs will often break out of the disk. During the expansion, some shell fragments will be accelerated in the plane of the gas disk, while others will be accelerated vertically, out of the plane of the disk. In this paper we are interested in the latter, which we will assume cover a fraction $C_\Omega$ of the sky as seen from the center of the cluster. Before any supernovae erupt in the driving cluster and before the shell fragments rise above the disk and become exposed to radiation or any supernova-driven hot outflows from other stars or clusters in the disk, their equation of motion is given by
\begin{eqnarray}
\label{eq: momentum disk1}
\frac{\mathrm{d} P_{\rm gas}}{\mathrm{d} t}&\simeq& F_{\rm rad} - F_{\rm grav}\nonumber\\
&\simeq& C_\Omega \left(\frac{\Lcl (t)}{c}-\frac{G \Msh^2(r)}{2r^2}\right)\,,
\end{eqnarray}
where $\Lcl$ is the luminosity of the cluster.  We argue that the breakout of shell fragments from the gas disk is the crucial step in driving a superwind. We refer to shell fragments that manage to rise above the gas disk as \emph{clouds}.
As we will show below, radiative driving, acting alone, can drive clouds above the disk scale height in a few Myr, i.e., before the explosion of any supernovae from the parent cluster.

\subsection{Scaling relations for ejecting gas clouds above the galactic disk}
\label{sec:scaling}

Here we provide an estimate of the critical gas and star formation rate surface densities required to launch a galactic wind, starting from the assumption that GMC shell fragment breakout is the limiting step. To do so we first estimate the mass and radius of the largest giant molecular clouds in a galaxy with a specified circular velocity $v_c$, disk radius $\Rd$ and gas surface density $\Sigma_g$. The surface density $\Sigma_G$ of the GMC then allows us to estimate the stellar mass of the largest star cluster in the GMC. Given the mass-radius relation for star clusters described in section \ref{sec:starclusters}, we can then compare the ejection velocity of the cluster to the galaxy circular velocity $v_c$.

We assume that galactic disks initially fragment on the disk scale height $H\simeq(v_T/v_c)R_d$ (given by hydrostatic equilibrium), where $v_T$ is the larger of the turbulent velocity or the sound speed of the gas. The fragments will form gravitationally bound structures with a mass given by the Toomre mass, 
\be \label{eq:Toomre}
\Mg=\pi H^2\Sigma_g\,.
\ee
The value of $v_T$ required to estimate $H$ can be obtained by assuming that the disk is marginally gravitationally stable, so that Toomre's $Q$ is of order unity: 
\be 
Q={v_cv_T\over \pi G R_d \Sigma_g}\sim 1. 
\ee
It follows that $H=\pi G Q (R_d/v_c)^2\Sigma_g$. The mass of a large GMC is then
\be \label{eq: GMC mass}
\Mg = \pi^3G^2Q^2\left({R_d\over v_c}\right)^4\Sigma_g^3
\ee
The star formation efficiency in a GMC, i.e. the limit beyond which the GMC is disrupted by radiation pressure, (see Eq.~\ref{eq:eps_g}), can be expressed as
\be
\epsilon_G = {\pi G c\over 2(L/M_*)} \Sigma_G={\pi G c\over 2(L/M_*)} \phi^2 \Sigma_g,
\ee
where $\phi\equiv H/\Rg\approx 2-3$, and the light-to-mass ratio $(L/M_*)$ is $3000\cm^2\s^{-3}$ for a \citet{2005ASSL..327...41C} initial mass function. The characteristic cluster mass is then 
\be \label{eqn: cluster mass}
\mcl=\epsilon_G\,M_g = {\pi^4G^3c\over2(L/M_*)} Q^2\phi^2\left({R_d\over v_c}\right)^4\Sigma_g^4.
\ee
The cluster ejection velocity, as given by eq.~\ref{eq:vc}, reads
\be
\vcl\simeq v_{esc}={ {\pi^2G^2c^{1/2}\over(2\,L/M_*)^{1/2}} {Q \phi\over \sqrt\rcl}
\left({R_d\,\Sigma_g\over v_c}\right)^2 }.
\ee
If this velocity is comparable to or larger than $v_c$, then a galactic scale wind will result. We define the velocity ratio
\be 
\label{eq:crit}
\zeta\equiv {\vcl\over v_c} ={\pi^2G^2c^{1/2}\over (2\,L/M_*)^{1/2}}\,{Q \phi\over \sqrt\rcl}\,{R_d^2\,\Sigma_g^2\over v_c^3} .
\ee
For $\zeta\gtrsim1$ we expect the largest star clusters in a galaxy to launch galactic scale winds. For a specified disk radius and circular velocity, this defines a critical gas surface density in order to launch a galactic wind.

\subsubsection{Critical star formation rate}
\label{sec:crit}

We now use (\ref{eq:crit}), in conjunction with the \citet{1998ApJ...498..541K} star formation law
\be
\dot\Sigma_* = \eta \left({v_c\over R_d}\right)\Sigma_g,
\ee
with $\eta=0.017$, to find the critical surface density of star formation required to launch a supergalactic wind. We find
\be
\dot\Sigma^{crit}_*={ (2\,L/M_*)^{1/4}\over \pi G c^{1/4}} \;{\eta\,\zeta\over\sqrt{Q\phi}}\;
{\rcl^{1/4}v_c^{5/2}\over R_d^2}.
\ee
Scaling to an $L_*$ galaxy,
\be
\dot\Sigma^{crit}_*\approx 0.06 \left({v_c\over200\kms}\right)^{5/2} \left({5\kpc\over R_d}\right)^2 M_\odot\yr^{-1}\kpc^{-1}.
\ee
From observations of galaxies with and without superwinds, \cite{2002ASPC..254..292H} gives $\dot\Sigma^{crit}_*\approx0.1\,M_\odot\yr^{-1}\kpc^{-2}$. 
Finally, we can express our radiatively-motivated threshold in terms of critical star formation rate
\be
\dot M_*^{crit}\approx 5\left(v_c\over200\kms\right)^{5/2}M_\odot \yr^{-1}.
\ee
%

\section{MODELING GALACTIC WINDS} \label{sec:galacticwinds}

We now extend the model used by \citet{2010ApJ...709..191M} to describe not only the destruction of a cluster's natal GMC but also the late-time evolution of a shell fragment that emerges from the gas disk. A number of new effects must be included: expansion and varying optical depth of the cloud, the radiation pressure contribution from neighboring clusters, ram pressure from supernovae, and the underlying dark matter potential. We present a dynamical model including these ingredients and show that radiation pressure from a massive star cluster expels cold gas above the host galaxy disk, where a combination of radiation pressure and ram pressure from a hot supernova driven wind then pushes the clouds out to scales reaching several tens of kiloparsecs.

Below we study the dynamics and properties of the gas as it travels above the disk. Its equation of motion is described by
\begin{eqnarray}
\label{eq: momentum disk2}
\frac{\d P_{\rm gas}}{\d t}&\simeq& F_{\rm rad} + F_{\rm ram} - F_{\rm grav}
\end{eqnarray}
where the radiation and ram pressure forces $F_{rad}(r,t)$ and $F_{rad}(r,t)$ include contributions from the parent cluster as well as other UV sources distributed over the disk and $F_{grav}$ describes the gravitational effect of the galaxy and its surrounding dark matter distribution. For simplicity we do not include the interactions any hot corona that might occupy the galaxy halo. We discuss its potential impact in section~\ref{sec:corona}.

\subsection{Launching the gas through a galactic halo}
We have shown that radiation from a single massive star cluster can eject clouds from the disk of a galaxy. However, neither radiation pressure nor ram pressure from supernova-driven hot winds arising from a single isolated cluster can drive the clouds to tens of kiloparsecs; doing so requires the collective effect of all the clusters in the galaxy, as we now show. 

\subsubsection{Radiative force}
Once the shell emerges from the galactic disk, it is subject to radiation pressure from stars in the disk. The observed surface brightness of star forming disks follows distributions that range from nearly constant for $R<\Rd$, where $R$ is the distance from the galactic center, to exponential, i.e., $\Sigma_{UV}\approx \exp^{-R/\Rd}$ \citep{2001ApJ...555..301M,2009A&A...501..119A}. For simplicity we will work with a constant surface brightness out to $\Rd$. If the height above the disk (which we approximate as $r-H$, where $r$ is the distance from the center of the star cluster) is smaller than $R_d$, the disk flux seen by the cloud, which quickly dominates that provided by the natal cluster, is roughly constant. The radiation pressure then becomes
\be
F_{\rm rad}= \left\{
    \begin{array}{ll}
        F_{\rm rad,cl} + \frac{L_{\rm gal}}{c}\,\left(\frac{r-H}{R_D} \right)^2 & \mbox{if } H<r<R_D\\
        F_{\rm rad,cl} + \frac{L_{\rm gal}}{c}\,\frac{1}{r^2} & \mbox{if } r>R_D\\
    \end{array}
\right.
\ee
where $r$ is the distance from the center of the cluster. At early times, the optically thick regime applies and the above relation does not depend on $\kappa$. Once the cloud reaches $r\gtrsim\Rd$, the flux seen by the cloud drops as $1/r^2$. 

The radiation pressure force also depends on the size of the cloud, which expands perpendicular to the direction of motion due to its finite temperature. The rate of expansion perpendicular to the radius vector will depend on a number of difficult to quantify factors, including the pressure of any hot component in the disk proper. The clouds are overpressured compared the ISM as a whole, since they are subject to the intense radiation from the central star cluster. If this pressure is large compared to the pressure of the hot gas component, which is the case while the cloud is in the gas disk, and possibly for heights of order the disk radius, the cloud will effectively expand into a vacuum; in this case the size of the cloud in the direction perpendicular to its motion will be given by $l_\perp=l_{\perp,0}+c_s(t-t_0)$, where $l_\perp$ is the radius of the cloud in the direction perpendicular to the radius vector, and $c_s$ refers to the sound speed of the cloud. 

If, on the other hand, the pressure of the hot gas in the disk is large, the cloud volume will vary with height above the disk, tracking the pressure in the hot gas. The latter will be in rough hydrostatic equilibrium, with a scale height much larger than that of the molecular gas disk, so the size of the cloud will not vary until it reaches a height of order the scale height of the hot gas. At that point the hot gas density will decrease as $1/r^2$, as will the density in the cloud. In most of the numerical work we present here we assume the cloud is overpressured compared to any hot gas; this is the case in the Milky Way, but the situation may be different in starburst galaxies.

If the cloud is at a distance less than $\Rd$ above the disk, any expansion in the perpendicular radius will give rise to an increase in the amount of radiation impinging on  the cloud. For $r>\Rd$, the $1/r^2$ radiative flux decrease will tend to cancel the effect of the increased area of the cloud, and the amount of light impinging on the cloud will increase less rapidly, or not at all.

The expansion of the cloud is also responsible for a third change in the radiative force; the cloud eventually becomes optically thin to the continuum emission from the disk. The fraction of the incident radiation that the cloud absorbs then drops like $1/r^2$, tracking the decrease in the column density and hence optical depth $\tau$. If $r>R_d$, the decrease in incident flux tends to cancel the geometric growth of the cloud, and the radiative driving, which is proportional to $\tau$, decreases as $r$ increases.

\subsubsection{Ram pressure from supernovae}

The hot gas from isolated supernovae is likely to be  trapped in the disk, so that it does not affect clouds above the disk. However, supernovae in large clusters will have a different fate. As we have just demonstrated, the ISM above large clusters will be expelled from the disk, opening up a pathway for the hot supernova gas to escape. 
We focus on the large scale behaviour of this hot gas, since the bulk of the cold gas is expelled from the vicinity of the cluster long before any stars in the cluster explode. We assume that the starburst lasts much longer then the lifetime of a very massive star (which we take to be $4\Myr$), so that we can average over many cluster lifetimes and use a mean supernova rate. 

Supernovae drive a hot wind. The ram pressure derived force on a cloud of cross-section $\pi l_\perp^2$ with a drag coefficient $C_D$ is
\be
F_{\rm ram} = C_D\pi l_\perp^2 \rho_h (v_h-v_c)^2\,.
\ee
The kinetic energy of the hot wind is given by
\be
 {1\over2}\dot M_h v_h^2 = \epsilon \,L_{\rm SN}.
\ee
where $\epsilon$ is the fraction of the supernova luminosity that is thermalized to produce a hot phase. It is generally found to range from 0.01 to 0.2 \citep{1992A&A...265..465T,1994MNRAS.271..781C,1997MNRAS.285..711P,1998ApJ...500...95T}. However, \citet{2009ApJ...697.2030S} find $\epsilon>0.3$, possibly reaching $\epsilon=1$, in the local starburst M82.

\begin{figure*}
\begin{center}
\includegraphics[width=.7\hsize]{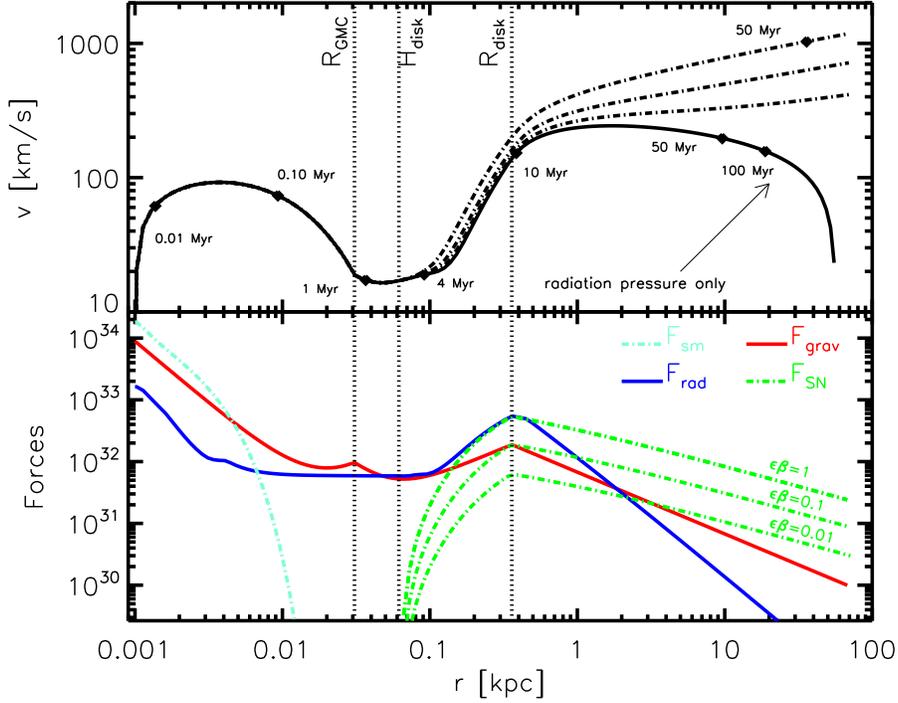}
\caption{The velocity (upper panel) and forces in a model of a $10^6M_\odot$ cluster embedded in a marginally stable ($Q=1$) starforming disk, plotted as a function of distance from the cluster center. The time since cluster formation is marked on the velocity curve in the upper panel. The forces shown in the lower panel  are; the force of gravity (red dashed line) due to the star cluster, the self gravity of the gas, and the gravity of the galactic halo, the force due to protostellar jets and possibly shocked stellar winds (cyan dashed line), which is confined to small scales, the radiation pressure force (solid blue line) and the ram pressure force from a supernova-driven hot outflow (sold green line). The radiation pressure force rises as the cloud emerges from the disk and is exposed to radiation from other cluster and stars in the disk; similarly the ram pressure force rises as the cloud enters the large scale hot outflow. In both panels, the vertical dotted lines mark the radius of the parent GMC, the scale height of the galactic disk, and the e-folding size of the galactic stellar disk (from left to right). }
\label{fig:cloud_velocity}
\end{center}
\end{figure*}

Following \citet{2009ApJ...697.2030S}, we introduce the mass loading parameter $\beta$
\be
\hot = \beta \dot M_{SN+SW}=0.2\beta\sfr,
\ee
where, by definition, $\beta>1$. The factor $0.2$ in the second equality corresponds to a \citet{2005ASSL..327...41C} initial mass function. For very high values of the mass loading $\beta$, the hot gas becomes radiative, cooling on a dynamical time \citep{2004ApJ...610..226S,2009ApJ...697.2030S}. This effectively limits the range of mass loading, $1\le\beta\lesssim17$. 

Using the above relations and considering that $v_c \ll v_h$, the ram pressure force can be expressed as
\be
F_{\rm ram}\simeq\left({\pi l_\perp^2\over 4\pi r^2}\right)0.2\,C_D\,\sfr\,\times\,\sqrt{\epsilon\,\beta}
\ee
The radiation pressure on the same cloud is
\be
F_{rad}= {\rc^2\over 4 R^2}\,\times\,\rm{min(\tau_{UV},1)}\,{L\over c}
\ee
The ratio of the two forces is
\be \label{eqn:force_ratio}
{F_{\rm ram}\over F_{\rm rad}} \approx {2\,C_D \sqrt{\epsilon\,\beta}}\,\times\,\frac{1}{ \rm{min}(1,\tau_{UV})}
\ee
For typical values: $C_D=0.5$, $\epsilon\approx0.3$, and $\beta=4$ and $\tau_{UV}=1$ in the optically thick regime, this ratio is $\sim1$. At large distances, of order $10\kpc$, when the clouds are optically thin, the ram pressure force will start to dominate.

\subsubsection{Gravitational force}
While in the disk, the shell takes part in the rotational motion of the galaxy, so that the force exerted by the galaxy on the GMC is cancelled by the centrifugal force due to rotation. However, as it rises above the disk, the cloud will lose the centrifugal support it enjoyed while pursuing its circular path around the galaxy. We take this into account by using
\be
g_{grav}= {v_c^2\over \Rd}\left(r\over R_d\right)
\ee
if $r<R_d$. This assumes that the GMC lies at a distance $\sim \Rd$ from the galactic center.

We model the overall gravitational effect of the galaxy and its dark matter distribution by using a singular isothermal sphere model for the galactic halo:
\be
g_{grav}={v_c^2 \over r}.
\ee
This accurately reflects observed rotation curves of spiral galaxies out to $r\approx50\kpc$, e.g., \cite{1991AJ....101.1231C}. It probably overestimates the force of gravity acting on the cloud at large radii (approaching the virial radius of the galaxy), where lensing measurements suggest that an NFW profile is a better fit, e.g., \citet{2006MNRAS.372..758M}.

\subsection{Numerical results}

Given the forces introduced above, we now solve the 1-D equation of motion and present results aimed at characterizing the fate of the shells described above. 

In Figure~\ref{fig:cloud_velocity} we consider the case of an M82-like galaxy, i.e. with a circular velocity $v_c=110\kms$ and a luminosity $L=5\times10^{10}L_\odot$ 
. For such a galaxy, our idealized model (Eq.~\ref{eq: GMC mass}) predicts a characteristic cluster mass of about $10^6\,M_\odot$, in agreement with observations \citep{2007ApJ...663..844M}.
As indicated in section~\ref{sec:scaling}, the $\zeta$ parameter of such a system is greater than one, indicating that the star cluster can drive a shell of gas above the plane. Our numerical calculation confirms the analytic result.
We now describe the detailed behavior of the shell as a function of time and radius. The bottom panel of the Figure presents the amplitude of the relevant forces as a function of radius:
\begin{itemize}
\item small scale force ($F_{sm}$, cyan dashed line): it represents the momentum per unit time deposited by proto-stellar jets, outflows and possibly shocked stellar winds on small scales. The details of this force are described in \citet{2010ApJ...709..191M}. Numerous observations of expanding bubbles in the Milky Way \citep{2010ApJ...709..424M} and in nearby galaxies \citep{2007AJ....133.1067W,2007ApJ...668..168G} demonstrate that such outflows start before any supernovae occur. The amplitude of this force quickly decreases, on a timescale of about 0.1 Myr.
\item Gravity (red dashed line): At early times the gravity is dominated by that exerted by the star cluster on the nascent shell; for $r<R_{GMC}$ this force drops like $1/r^2$. At larger radii, $H\lesssim r\lesssim R_d$, the mass in the shell is constant, but the cloud is losing the rotational support it enjoyed in the disk, so the effective gravitational force increases. At still larger radii the force is dominated by the interaction between the cloud and the combined gravity of the stars and dark matter in the galaxy proper, so that $g(r)=v_c^2/r$. 
\item Radiation pressure (blue solid line): initially, the radiation pressure force originates only from the parent star cluster, and the shell is optically thick to far infrared photons. As a result, this force initially scales as $1/r^2$, but the expansion of the cloud gradually reduces the far IR optical depth below unity, and the net radiation pressure force becomes roughly constant. The radiation pressure from the parent cluster lasts only for a period of about 8 Myr, i.e. the time for the cluster luminosity to decrease by a factor of two. If the shell does not reach an altitude greater than the height of the disk during that time, it will fall back toward the mid plane of the galaxy. If the cloud succeeds in rising above the disk, it then feels the radiation pressure from the neighboring clusters, which increases until $r=R_d$. Above this height, the radiative flux decreases like $1/r^2$. The fraction of the luminosity impinging on the cloud is given by the area of the cloud divided by $r^2$. On those scales, the size of the cloud increases roughly linearly with time, but the radius increases more rapidly. As a result, the momentum deposition decreases. Finally, due to its continuing  expansion, the cloud becomes optically thin at about half a $\kpc$, so $F_{rad}$ drops like $1/r^2$ beyond this point. 

\item Ram pressure (green solid line): we have already noted that observations of the Milky Way and the LMC show that stellar winds are not dynamically important, so the ram pressure force is gradually ramped up when the cool gas emerges from the disk, around $60\pc$ in figure \ref{fig:cloud_velocity}. It is comparable to the radiation pressure, with the exact ratio of the two forces varying with $\epsilon$ and $\beta$, as in eqn. (\ref{eqn:force_ratio}). The ratio changes when the cloud becomes optically thin to UV radiation, around $r\simeq0.5\kpc$. At large radii, the ram pressure is the dominant force, causing the cloud to accelerate slowly beyond $r\sim50\kpc$. Because the area of the cloud continues to increase with increasing $r$, but not as rapidly as $r^2$, the ram pressure force decreases with increasing $r$.

\end{itemize}

From an observational point of view, we note the following results: clouds escape the disk in less than $3\Myr$, before any supernovae explode in the cluster. They reach velocities of several hundred kilometers per second at a radius of $10\kpc$, after a time $\sim 10\Myr$. The clouds reach a distance of $50\kpc$ after $\sim100\Myr$.

\subsubsection{Star-forming galaxy: varying cluster mass}

\begin{figure}
\includegraphics[width=\hsize]{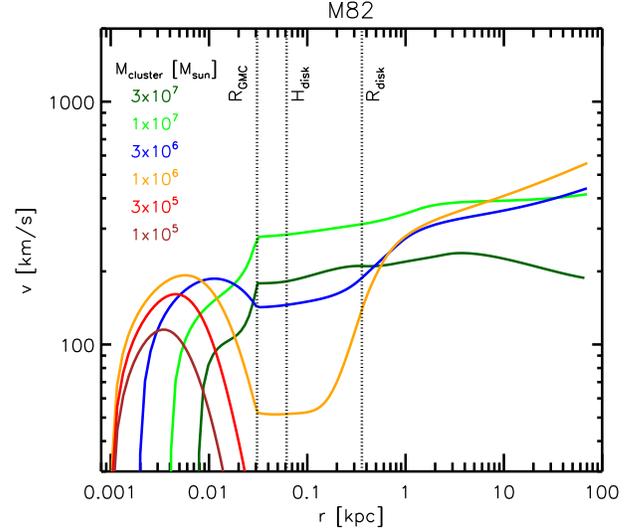}
\caption{Outflows launched by clusters in a M82-like galaxy. The different colors denote clusters with different masses, ranging from $10^5$ to $3\times10^7M_\odot$. We note that clusters with $M<10^6\,M_\odot$ only trigger galactic fountains.}
\label{fig:M82_varying}
\vspace{.2cm}
\end{figure}

As shown in Eq.~\ref{eq:crit}, for a given galaxy the main parameter defining the fate of gas clouds ejected by radiation pressure is the cluster mass---only sufficiently massive clusters can transfer enough momentum to eject a shell of gas. To demonstrate this result numerically, we use the M82-like galaxy parameters described above, and solve the equation of motion for a range of cluster masses, from $10^5M_\odot$ up to $10^8M_\odot$. The results are presented in Figure~\ref{fig:M82_varying}. We can see that, in a halo of mass $M_h\approx10^{11} M_\odot$, clusters with $M=10^5M_\odot$ will launch galactic fountains rather than galactic winds. More massive clusters launch winds to larger distances from the center of the galaxy, with clusters above $M=10^7M_\odot$ expelling gas from the halo. 

\subsubsection{Quiescent galaxies: varying cluster mass}

\begin{figure}
\includegraphics[width=\hsize]{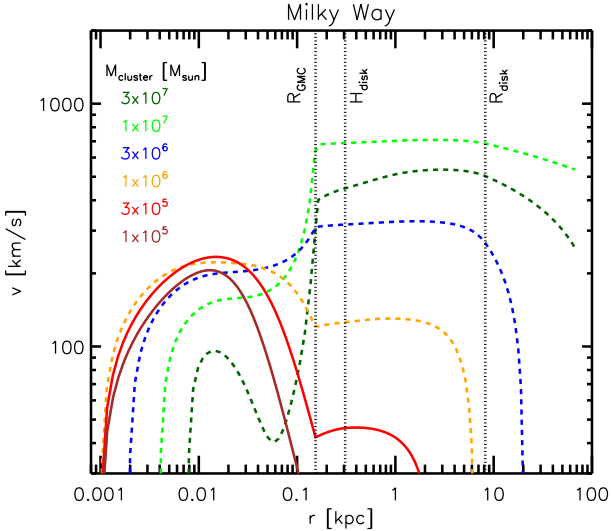}
\caption{Outflows launched by clusters in a Milky Way like galaxy. The dashed lines represent the trajectories of material launched by clusters with a mass exceeding that allowed by the Toomre criterion. Our model shows that such a galaxy can only trigger galactic fountains, in agreement with the scaling relation introduced in Eq.~\ref{eq:crit}.}
\label{fig:MW_varying}
\end{figure}

We now consider the case of a quiescent galaxy: The Milky Way.
As done above, we study the evolution of the material surrounding young clusters. 
In such a system, $v_c=220\kms$, $\M_g=2\times10^9\,M_\odot$ and $R_d=8\,\kpc$.  The results are shown in Figure \ref{fig:MW_varying}. 
As a result of the Toomre criterion (see Eq.~\ref{eq: GMC mass}), typical star clusters are expected to have $\mcl=10^4-10^5M_\odot$.
The trajectories of shell fragments expelled by such star clusters are shown in solid line. Those of hypothetically more massive clusters are represented with dashed lines. As can be seen, a galaxy like the Milky Way is able to launch galactic fountains to heights of order a kiloparsec. However, even clusters with $M_*=10^5 M_\odot$, similar to the most massive clusters in the Milky Way, do not drive large scale outflows. Such a behavior is in agreement with the scaling relation presented in Eq.~\ref{eq:crit}. It reflects the fact that in such a galaxy, the initial kick resulting from radiation pressure is not strong enough to trigger a large-scale outflow.

\subsubsection{Starburst galaxies: varying cloud mass}

We now model the evolution of shells of various masses in a typical $z=2$ starburst galaxy. We parametrize such a system with $v_c = 100\kms$, $\M_g=2\times10^{10}\,M_\odot$, $R_d=5\,\kpc$ and $L=4\times10^{11}\,L_\odot$. We consider the case of a Lyman break galaxy for which most of the luminosity comes out in the optical. Given Eq.~\ref{eq: GMC mass}, the star cluster mass expected for such a galaxy is about $3\times10^8\,M_\odot$. This stellar mass will be distributed in a number of clumps, reducing the self gravity. To take this effect into account, we increase the timescale over which the radiation pressure from the parent cluster is on by a factor four.

Here, as an illustration, we use do not include the ram pressure contribution from supernovae and consider only radiation pressure . We find that, for such a galaxy, radiation pressure can expel the entire shell with a typical velocity $v\simeq100\kms$. We also consider the case of lower mass shell (or shell fragments): these can easily 
reach velocities of order $1\,000 \kms$ and might be related to the tails of blue-shifted self absorption observed in star forming galaxies (Weiner et al. 2009, Steidel et al. 2010).





\begin{figure}
\includegraphics[width=\hsize]{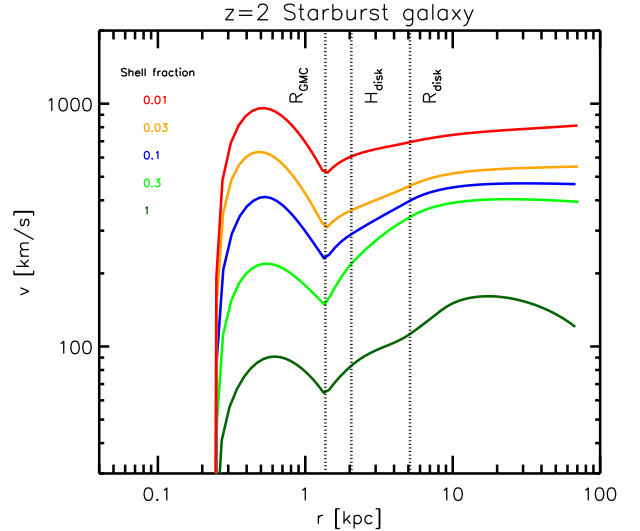}
\caption{Outflows launched by a $z\sim2$ starburst galaxy. Small shell fragments can reached outflow velocities of order $1\,000 \kms$.}
\vspace{.2cm}
\label{fig:varying_sizes}
\end{figure}

\section{CLOUD AND GLOBAL WIND PROPERTIES} \label{sec:properties}

We have shown that radiation pressure from a single massive cluster can drive cool gas above the disk of a galaxy. Subsequently, the radiation and supernova supplied ram pressure from the collection of star clusters in the disk, can drive the cool gas to radii of tens or even hundreds of kiloparsecs. This outflow will be in photoionization equilibrium, with a temperature of order $10,000\K$ and an ionization parameter of order $U\sim10^{-4}$. It will be dusty, and it will have the bulk of its ions in a moderately low ionization state. The columns seen in absorption will range from $N_H\sim10^{22}\cm^{-2}$ downward, decreasing with increasing radius and velocity. The cover factor $C_f$ of the individual outflows will also increase with increasing radius and velocity. 

The column densities of the clouds that are launched are initially those of the parent GMCs, i.e., $N_H\sim10^{22}\cm^{-2}$ and upwards. The column decreases roughly as $\Sigma_G(\Rg/r)^2$ once the shell has disrupted the parent GMC, an approximate relation that holds exactly if the radial velocity is constant at large radii.

The bulk of the gas in starburst galaxies is in molecular clouds (unlike quiescent galaxies such as the Milky Way, which has a substantial mass of atomic gas). It follows that most of the mass in the bubble shell (and later in the shell fragments in the wind) originates in the natal GMC. However, it is probably true that some gas lies either above or below the GMCs in which star formation occurs. This overlying gas will slow or possibly halt the bubble shell, if it has a column similar to that of the GMC. However, we note that star clusters in the Milky Way are seen to have a dispersion in their distance from the disk midplane that is comparable to the gas scale height $H$, e.g., \citep{2010ApJ...709..424M}. As a result it seems likely that a substantial fraction of clusters above the critical cluster mass will drive winds out of one side of the disk.

It seems likely to us that clusters in particularly large GMCs, with GMC radii comparable to $H$ such as those seen in chain or clump galaxies, can expel winds from both sides of the galactic disk. Our one dimensional models cannot address this question without further information on the distribution of gas outside GMCs. More detailed modelling will be needed to decide this question.

\subsection{Cloud properties}

If a wind is launched from a cluster, the column density through the wind will be given roughly by
\be \label{eq: column}
N_H(r) \approx N_g
\left({H\over r}\right)^\beta,
\ee 
where $N_g\equiv\Sigma_g/m_p$ and $\beta\approx1-2$. The power law index $\beta$ depends on the details of the model, in particular on the initial size of the cloud, how $v(r)$ varies, and on whether the cloud is pressure confined by hot gas. The initial column density is simply that of the disk (when the GMC has expanded to the size of the disk scale height). As already noted, the gas will be in photoionization equilibrium, and illuminated by thousands of O stars, so its temperature will be of order $10^4\K$. Since the shell or cloud is being accelerated, the width of the shell in the direction of the acceleration will be
\be l_{||} \approx {c_s^2\over g}, \ee where the acceleration $g\approx v^2/r$. Since $v\approx v_{cl}\approx v_c$, we have
\be \label{eq: parallel}
l_{||} \approx \left({c_s\over v_c}\right)^2 r\approx8\times10^{-3}r
\left({T\over 10^4\K}\right)
\left({100\kms\over v_c}\right)^2
\ee
Combining equations (\ref{eq: column}) and (\ref{eq: parallel}), the number density is
\begin{eqnarray}
n(r)&\approx&{N_g\over r}
\left({H\over r}\right)^\beta
\left({v_c\over c_s}\right)^2\hfill\nonumber\\
&\approx& 250\left({N_g\over 10^{23}\cm^{-2}}\right)
\left({H\over 50\pc}\right)^2
\left({1\kpc\over r}\right)^3\cm^{-3},\hfill
\end{eqnarray}
where we have taken $\beta=2$ and $v_c=100\kms$. For a star formation rate of $5M_\odot\yr^{-1}$, this corresponds to an ionization parameter
\be
U\equiv{Q\over 4\pi r^2 n c}\approx 10^{-3}
\left({Q\over 10^{54}\s^{-1}}\right)
\left({r\over 1\kpc}\right),
\ee
where $Q$ is the number of ionizing photons emitted by the galaxy per second; we have scaled to a value appropriate to the assumed star formation rate. If $\beta<2$, the $r$ dependence is weaker, e.g. $\beta=1$ implies $U=const.$ We have run the photoionization code CLOUDY \citep{1998PASP..110..761F} to determine the ionization state of the gas. The bulk of magnesium is in the form of MgII for $U=10^{-3}$, while a fraction of order $0.1$ of sodium is neutral. At very large $r$ the background UV flux will dominate that from the galaxy, and the ionization fractions of both ions will drop, even for the case $\beta=1$. 

The column densities, densities, sizes, and ionization parameters just given are those of individual cloud in the outflow from a single cluster. Observations of expanding gas around clusters in the Milky Way make it clear that while shell-like structures do form, they are not very regular, nor is all the material at the same radius from the cluster. Physically, it is clear that higher column density gas will accelerate less rapidly than lower column density gas, leading to clouds at a range of radii. Further, it is likely that material at the fringes of the clouds will be ablated, as is seen in star forming regions in the Milky Way. When seen in the large, the outflows from a cluster will appear quasi-continuous; if the flows from multiple clusters cannot be resolved, the outflow from the galaxy will appear even more continuous.

\subsection{Absorption covering factor}

We note that the outflow will not be seen as such until the continuum (dust) optical depth drops below unity; only then may absorption lines be seen against the cluster or galactic light. The corresponding column is $N_H\approx10^{21}\cm^{-2}$, depending on the wavelength of the absorption line used to trace the outflow. 

While the continuum optical depth drops below unity at fairly small radii, the MgII resonance transition at $2900$ \AA\ is optically thick out to radii as large as $100\kpc$. The dynamical time to 100kpc is $\sim 100\Myr$. Hence we expect blue shifted absorption to be seen in post-starburst galaxies as well as in active starbursts. 

Individual GMCs and their cluster progeny occupy a small fraction of the disk area, so individual outflows have a small disk covering factor
\be
C_{fd}\equiv {\rc^2(v)\over \Rd^2}.
\ee
This covering factor is an increasing function of cloud velocity $v$ for small distances $r$, since initially both $v(r)$ and $\rc(r)$ are increasing functions of $r$. When the clumps enter the large scale hot outflow, their radii may decrease as they come into pressure equilibrium, but after that they will resume their expansion. However, the summed cover factor of all the clouds will be a significant fraction of unity.

The ratio of disk scale height to disk radius is related to the star formation rate of the galaxy. For high star formation rates, it can reach $(H/\Rd)^2\approx0.1-0.3$, but is smaller for most starburst galaxies. However, clusters above and below the disk midplane can drive gas from either side of the disk (and possibly from both sides, as just noted), so that sightlines from most directions will encounter the outflow from one or more clusters.

\subsection{Mass loss rates}
\citet{2010ApJ...709..191M} show that the fraction of gas in a GMC that ends up in
stars is given roughly by
\be
\eg\approx {\Sigma_G\over 1+\Sigma_G/0.25\g\cm^{-2}}\,,
\ee
interpolating between their optically thin and optically thick expressions. The balance of the GMC is either returned to the interstellar medium or driven out of the galaxy as a wind. We have argued in this paper that for systems above the threshold given in \S \ref{sec:crit}, one quarter to one half of the GMC is launched as a wind. The ratio of mass launched out of the disk to mass locked up in stars is very roughly
\be
{\dot M_w\over \dot M_*}\approx{0.25\g\cm^{-2}+\Sigma_G\over\Sigma_G}.
\ee
This predicts a very large ratio of mass lost in a wind compared to stellar mass in low $\Sigma_G$ systems. Since we require that the galaxy be above the threshold star formation rate, this high mass loss ratio applies to galaxies with low circular velocities. 

In addition to this local criterion, there is a global limit to the mass loss rate driven by radiation pressure, namely that the rate of momentum carried out by the wind is no larger than the rate of momentum supplied by starlight. This limit is
\bea
\dot M_w &=& \left(1+{F_{\rm ram}\over F_{\rm rad}}\right)
L/c v_\infty\\
&=& \left(1+{F_{\rm ram}\over F_{\rm rad}}\right)
\left[
\left({L\over M_{cl}}\right)
{\tau_{ms}\over c^2 }
\right]
\left({c\over v_\infty}\right)
\dot M_*.
\eea
The maximum wind mass loss rate per star formation rate for such a radiatively driven outflow is
\be
{\dot M_w \over \dot M_*} \approx0.4\left(1+{F_{\rm ram}\over F_{\rm rad}}\right)
\left({v_\infty\over 300\kms}\right)^{-1}.
\ee

These expressions assume that the bulk of the star formation occurs in clusters above the critical mass necessary to launch galactic winds. As noted in section~\ref{sec:mass function}, the mass function of star clusters is shallow so that this criterion will be satisfied for galaxies with sufficiently high star formation rates.

\section{DISCUSSION} \label{sec:discussion}

\subsection{Comparison to other launching mechanisms in star clusters}


\subsubsection{Ram pressure from supernovae and cloud survival}

There is little doubt that hot gas from supernovae pushes cool gas out of galaxies.
However, it has been argued here and elsewhere that supernovae do not push cool gas out of GMCs. There is good evidence in the Milky Way that GMCs are disrupted well before any supernovae explode in them \citet{2009ApJ...693.1696H,2010ApJ...709..424M}. 

The crucial point, however, is that cool gas entrained in a hot flow on GMC scales will not survive at a low temperature long enough to escape the galaxy. As noted in the introduction, clouds caught up in a hot outflow are compressed on a cloud crossing time, $t_{cc}=\chi^{1/2}\rc/v_b$. For a cloud with $T=1000\K$ at a distance $r\approx10\pc$ from its natal cluster, we find 
\be
t_{cc}\approx6\times10^{12}
\left({\rc\over 1\pc}\right)\s,
\ee
where we have scaled to the temperature $T_h=5\times10^7\K$ of the hot flow seen in M82 \citep{2009ApJ...697.2030S}.

Numerous simulations of interactions of hot density gas with cool ($T\le10^4\K$) gas demonstrate that the clouds are rapidly destroyed, i.e., in $\sim 3-5t_{cc}$,  in the simulations of  \citet{1994ApJ...420..213K} and \citet{2002ApJ...576..832P}. These early simulations employed an adiabatic equation of state, so that the temperature of the cool cloud increased as it was compressed. Simulations which include radiative cooling find that clouds live longer, $\sim 10t_{cc}$, reaching velocities of several hundred kilometers per second  \citep{2002A&A...395L..13M,2009ApJ...703..330C,2010arXiv1002.2091P}. For example, \citet{2009ApJ...703..330C} employ high resolution 3D radiative models, finding that a substantial fraction of the gas in a cloud is lost, via Kelvin-Helmholtz and Rayleigh-Taylor instabilities, to the hot flow by the time the cloud has traveled $75\pc$ (about a million years), consistent with the simple analytic estimate of $\sim 10t_{cc}$. 

The short lifetimes of such ram pressure driven clouds was stressed by \citet{2005MNRAS.362..626M}, who noted another form of destruction, that of thermal conduction. They find that conductive heating suppresses the Kelvin-Helmholtz and Rayleigh-Taylor instabilities that disrupt the clouds in non-conductive simulations. However, the conduction also causes rapid mass loss from the clouds, leading to lifetimes $\sim1-10\Myr$. As in the simulations of \citet{2008ApJ...674..157C}, the clouds are essentially completely disrupted by the time the reach $r\approx1\kpc$, even in the most favorable simulations. 

Observations of the solar wind have shown that the conductivity is similar to the classical Spitzer value, e.g., \citet{2003ApJ...585.1147S}, modified to account for saturation at the free-streaming heat flux. This, combined with the rough agreement between the analytically estimated cloud lifetimes against conduction and those found by the numerical simulations of \citet{2005MNRAS.362..626M}, strongly suggest that ram pressure driven cold gas clouds have lifetimes of order $\sim1\Myr$, simply due to heat conduction, independent of the effects of Kelvin-Helmholtz instabilities.

To summarize, the destruction time $\td\lesssim20\tcc$ of cold clouds entrained in a hot outflow is given by
\be
\td\lesssim 
4  
\left({T_h\over 5\times10^7\K}\right)
\left({T_c\over 10^3\K}\right)^{-1}
\left({a\over 1\pc}\right)~{\rm Myr}\,.
\ee
Assuming a cloud velocity of $300\kms$, the cloud is destroyed after traveling a distance
\be
R_{\rm destruction}\lesssim 1\kpc
\left({a\over 1\pc}\right).
\ee
This is a rather generous estimate, since it assumes the upper limit for estimates of the cloud destruction time, and assumes that the clouds are accelerated instantaneously. 
The sizes of starburst disks range from $R_d\approx 0.3-1\kpc$ or larger, implying that the hot gas density does not begin to drop until $r=0.3-1\kpc$. 

We conclude that cold clouds entrained in hot flows on GMC scales do not survive beyond distances of $\sim 1\kpc$. 

Radiatively driven cold gas clouds escape their natal clusters well before any supernovae explode, so they are not subject to the ill effects of hot gas, at least until they reach the surface of their galactic disk. If the hot gas fills the volume above the disk surface, the cold clouds will be subject to both destruction by Kelvin-Helmholtz instabilities and conductive evaporation, but at a much lower rate than the clouds simulated above. The slower evaporation results from the fact that the conductivity is saturated; in that case the mass loss rate $\dot M_{\rm cond}\sim n_h^{5/8}$, where $n_h$ is the number density of the hot gas. The longer Kelvin-Helmholtz times follow from the larger length scale $\rc$ of the clouds, which grows as the clouds move away from the launching cluster.

This leads to a clear observational distinction between the two types of driving. Small clouds driven by hot gas have lifetimes of order $1\Myr$ or shorter in most simulations; in the most optimistic case the lifetime may reach $\sim3\Myr$. The cold clouds survive to $r\lesssim 1\kpc$ from the disk (assuming a rather generous instantaneous  $v=300\kms$). 

In contrast, radiatively driven clouds are not subject to such rapid initial mass loss. They will travel for $\sim4\Myr$, reaching distances $300\pc$ before the hot gas from the first cluster supernovae explode, and about twice that before the bulk of the supernovae in the cluster explode. The cold clouds will be traveling at $\sim100\kms$ or more when the hot gas reaches them, and they will have sizes of order $\sim 50\pc$, assuming only that they expand at their sound speed (initially the clouds may expand more rapidly, as the radiation field diverges). As a result, radiatively driven clouds will survive to reach $r\sim50\kpc$ or more. Cool outflows extending to $r\gg1\kpc$ from the host disk are a clear signature of a radiatively driven outflow.

A second test that can distinguish between radiative driving and supernova driving is even more direct. Individual star clusters are known to drive winds. If radiative driving is important, winds will be seen emanating from clusters younger than $4\Myr$. Since no stars have lifetimes shorter than this, supernova driving cannot account for any winds seen emerging from such young clusters. 

\subsubsection{Shocked stellar winds}

Other mechanisms have been proposed to drive bubbles in the Milky Way and winds in other nearby galaxies. For example, shocked stellar winds from O stars produce $10^6$K to $10^7$ K gas seen via X-ray emission. This gas was suggested to be the driving force behind the bubbles observed in the Milky Way by \cite{1975ApJ...200L.107C}. However, on closer examination, these models fail to reproduce the observations. The predicted bubble radii are much larger than observed, e.g. \citep{1996ApJ...465..231O}, as is the diffuse X-ray emission from the shocked gas. Thus the shocked stellar winds either cool in place without emitting X-rays \citep{1984ApJ...278L.115M} or escape from the bubble through holes in the bubble wall \citep{2009ApJ...693.1696H}. In either case they do not affect the dynamics of the bubble. 

\subsubsection{Other hybrid models}

Both \citet{2005ApJ...618..569M} and \citet{2009MNRAS.396L..90N} discuss a combination of radiative and ram pressure driving for galactic superwinds, but neither paper addresses the fact that most stars form in clusters, nor do they mention the short lifetimes of ram pressure driven outflows originating at small radii. The latter implies that cold outflows are restricted to a few kiloparsecs around the host galaxy, in contrast to the prediction made here, that cold gas survives to $10-100\kpc$. 

We find that clusters are crucial for launching winds; the radiation from massive clusters blows holes in the local ISM, starting the cold gas on its way, and allowing for the hot gas from subsequent supernovae to escape readily. This hot gas can then drive cold gas, both that from its own natal cluster, and cold gas from other, older clusters to large radii. 

We have stressed that launching cold gas above the disk is the crucial step in driving a superwind. We used this point to estimate the critical star formation rate surface density. Because the radiative force is less destructive than ram pressure driving by hot gas, and because the clouds grow in size as they move away from their launching points, the clouds can survive the any subsequent blast from hot gas.

\subsection{The hot corona}
\label{sec:corona}

We have neglected any hot gas which might reside in the galaxy halo, and which could potentially halt the outflow of cold gas clouds within the halo. This is drag (or destruction) is relevant for any proposed launching mechanism.

If the halo contained the cosmic abundance of baryons, $M_b\approx0.173M_h$ \citep{2009ApJS..180..306D}, the mean column of hot gas would be $N_H\sim3\times10^{19}\cm^{-2}$. Since this is much larger than the cloud column density at the virial radius, the hot gas would either stop or shred the outgoing clouds. However, the hot gas content of cluster and group size halos has been measured to be smaller than the cosmic abundance, e.g., \citet{2009arXiv0911.2230D}. These authors measure a hot gas fraction $f_g$ that ranges from $f_g\approx0.08\pm0.02$ for galaxy halo masses $M_h=6\times10^{14}M_\odot$ down to $f_g=0.013\pm0.005$ for $M_h=2.6\times10^{13}M_\odot$. Written as a ratio to the cosmic baryon abundance, this is $f_g/f_c=0.46$ to $0.075$. It is believed that the hot gas fraction in galaxies ($M_h\lesssim10^{13}M_\odot$) is yet smaller \citep{2007ARA&A..45..221B}. If we take $f_g/f_c=0.1$, then cold clouds reach $r\sim 50-100\kpc$ before they run into their own column of hot halo gas.

\section{CONCLUSIONS} \label{sec:conclusions}

We have shown that radiation pressure from massive stellar clusters can drive outflows with velocities $v\sim{\rm a~few}\times 100\,\kms$ to distances $\sim50-100\kpc$ from their host galaxies. The outflows consist of cold ($\sim10^4\,\K$) and dusty clouds with characteristic column densities in the range $N_H\sim10^{21}\cm^{-2}$ near the galactic disk, ranging down to $N_H\sim10^{18}\cm^{-2}$ at $r\sim 50\kpc$. The gas in the clouds is in a low ionization state. 
Once a cloud is above the disk of a galaxy, it experiences the radiation pressure from nearby clusters as well as ram pressure from supernovae.

This solves a major problem of hot outflow-driven cold clouds, namely their short lifetimes. Since radiatively driven clouds need not be in contact with (relatively) dense hot gas in order to be driven out of the galaxy, they are not destroyed by
Kelvin-Helmholtz and/or conduction.

We have argued that individual clusters can drive outflows above the disk. If the disk is sufficiently luminous, the disk light can then drive the outflow to distances of tens of kiloparsecs. This evades the need for a global ``blowout'' of the interstellar medium of (the central portions of) the galaxy in order to drive a large scale wind. We note that ULIRG are clearly not ``blown-out'' and yet have winds, so this is an attractive feature of cluster driven winds.

Since the gas is launched from individual star clusters roughly vertically from the disk, it will retain the rotational motion of the disk. If the launching cluster is at large disk radius, measurements of the rotational velocity will find gas at large radii and large azimuthal velocities. The mass loss rates from cluster driven winds are comparable to the star formation rate in massive galaxies, and substantially larger than the star formation rate in low mass galaxies. In both types of objects, the terminal velocity of the wind is similar to the rotational velocity of the galaxy. While the winds may or may not reach the escape velocity, they can nevertheless reach radii in excess of the host galaxy's virial radius.

Since the cluster escape velocity must be comparable to the escape velocity from the galaxy to launch a flow above the disk, and because the maximum cluster mass is related to the gas surface density and hence star formation rate, there is a critical star formation rate (per unit area) $\dot \Sigma_*\approx0.1\,M_\odot\yr^{-1}\kpc^{-2}$ to launch a wind radiatively. This is in good agreement with the threshold inferred from observations. 

This paper has employed simply analytic and one dimensional hydrodynamic models of cluster driven winds to argue that such winds are crucial to the formation of galactic superwinds. The results provide strong motivation to carry out more realistic three dimensional radiative hydrodynamic modeling.

\acknowledgments

N.M.~is supported in part by the Canada Research Chair program and by
NSERC. T.A.T.~ is supported in part by
an Alfred P.~Sloan Fellowship.

\bibliography{Cluster_wind}

\end{document}